\documentstyle[12pt,aaspp4]{article}

\def\ab{$\sim$}

\def\etal {{et~al.}}

\def\HI {H\kern0.1em{\sc i}}
\font\caps=cmcsc10 scaled 1200

\lefthead{put in names for page header}
\righthead{put in abbreviated title for page header}
\slugcomment{To appear in the Astrophysical Journal}

\begin{document}

\title{The COINS Sample -- VLBA Identifications of Compact Symmetric Objects } 

\author{A. B. Peck\altaffilmark{1,2,3} \&
G. B. Taylor\altaffilmark{1}} 
\altaffiltext{1} {National Radio
Astronomy Observatory, P.O. Box O, Socorro, NM
87801;\\apeck@nrao.edu,gtaylor@nrao.edu} \altaffiltext{2}{Physics
Dept., New Mexico Institute of Mining and Technology, Socorro, NM
87801} \altaffiltext{3}{Now at Max-Planck-Institut f\"ur
Radioastronomie, Auf dem H\"ugel 69, Bonn D-53121, Germany}
\setcounter{footnote}{0}

\begin{abstract}

We present results of multifrequency polarimetric VLBA observations of
34 compact radio sources.  The observations are part of a
large survey undertaken to identify CSOs Observed in the Northern Sky
(COINS). 
Compact Symmetric Objects (CSOs) are of particular interest in
the study of the physics and evolution of active galaxies.  
Based on VLBI continuum surveys of $\sim$2000 compact radio
sources, we have defined a sample of 52 CSOs and CSO candidates.  In
this paper, we identify 18 previously known CSOs, and introduce 33 new
CSO candidates. We present continuum images at several frequencies
and, where possible, images of the polarized flux density and spectral
index distributions for the 33 new candidates and one previously known
but unconfirmed source.  We find evidence to support the inclusion of
10 of these condidates into the class of CSOs.  Thirteen candidates,
including the previously unconfirmed source, have been ruled out.
Eleven sources require further investigation.  The addition of the 10
new confirmed CSOs increases the size of this class of objects by
50\%.

\end{abstract}
\keywords{galaxies:active -- quasars:general -- radio
continuum:galaxies -- surveys}

\clearpage
\section{Introduction}

Compact Symmetric Objects (CSOs) are compact ($<$1 kpc) sources with
lobe emission on both sides of an active core (Wilkinson et al.\ 1994).
The study of these objects is of particular interest because the small
size of the CSOs is thought to be attributable to the youth of the
source itself (ages 10$^3$ -- 10$^4$ yr), rather than its confinement
by a dense medium (Readhead et al.\ 1996a).  Unifying evolutionary
models have been proposed (Readhead et al.\ 1996b, Fanti et al.\ 1995)
whereby these CSOs evolve into Compact Steep Spectrum (CSS) sources,
and then into Fanaroff-Riley (1974) Type II objects.

Recent studies have shown that the majority of detections of \HI\ 
absorption in galaxies has been in CSOs and Steep Spectrum Core (SSC)
objects (Conway 1996; Peck \etal\ 1999), rather than core-dominated
radio sources.  Of the theories proposed to explain this difference,
the existence of a circumnuclear disk or torus structure seems the
most likely.  In this scenario, it is the orientation and geometry of
the sources which is the cause of the discrepancy.  The core-dominated
sources are oriented close to the line of sight and the jets are
comprised of extremely high velocity outflow. This causes the
approaching jet to be strongly Doppler boosted, while the counterjet
is Doppler dimmed.  In CSOs, on the other hand, most of the continuum
emission is not beamed, and thus the counterjet can contain up to half
of the flux density of the radio source.  Obscuration by a
circumnuclear torus can then be seen against this counterjet, and if
the structure has a significant fraction of atomic gas, \HI\ 
absorption can be expected (Conway \& Blanco 1995).  In some cases
(e.g. 1946+708 -- Peck, Taylor \& Conway 1999) free-free absorption
provides further evidence of a dense circumnuclear torus.  


Another benefit of studying sources oriented at small angles to the
plane of the sky is that in some cases jet velocities can be measured
for both the jet and counterjet.  Under the assumption that components
are ejected simultaneously from the central engine, observations of
proper motions can provide a direct measure of the distance to a
source, and constraints can be placed on the value of H$_0$ (Taylor \&
Vermeulen 1997).  


Unfortunately, only a few CSOs have been studied to date.  In these
few CSOs, the kinematics of the atomic hydrogen seen in absorption on
parsec scales is intriguing, but complicated.  It is clear that a more
comprehensive sample is required to fully understand the nature of the
circumnuclear gas in these objects.  The CSOs Observed in the Northern
Sky (COINS) sample, defined in \S2, is an attempt to identify a larger
sample of CSOs which can be comprehensively studied using VLBI
techniques.  Here we identify 18 previously know CSOs, listed in Table
\ref{tab1}.  \HI\ absorption studies toward several of these are
currently underway, with 4 yielding published detections.  Table
\ref{tab1}\ also lists 33 new and one previously known but unconfirmed,
CSO candidates.  In \S3 we describe multi-frequency VLBI polarimetric
follow-up observations of sources in the COINS sample.  Results are
presented in \S4 and discussed in \S5.  Future papers will address
bidirectional jet motions and free-free absorption in the COINS
sample.  Before \HI\ absorption studies can be undertaken, optical
spectroscopy is required to determine the redshifts of the sources. 
The current status of the source redshifts, and \HI\ absorption
studies, is given in Peck \etal\ (1999).

\section{Sample Selection Criteria}

The class of Compact Symmetric Objects are distinguished from other
compact radio sources by their high degree of symmetry.  An
explanation for this symmetry is that CSOs tend to lie close to the
plane of the sky.  For this reason, relativistic beaming plays a minor
role in their observed properties, resulting in little continuum
variability (Wilkinson \etal\ 1994). They usually have well-defined
lobes and edge-brightened hotspots on either side of an active core,
often exhibiting a striking "S" shaped symmetry (Taylor, Readhead,
\& Pearson 1996a).

Since CSOs are rare (\ab2\% of compact objects, see \S5), it is
necessary to start with large VLBI surveys and go to moderately low
flux density levels (\ab100 mJy at 5 GHz) in order to obtain the 52
CSO candidates that make up the COINS sample.  The sources in the
COINS sample have been identified based on images in the
Pearson-Readhead (PR; Pearson \& Readhead 1988), Caltech-Jodrell Bank
(CJ; Polatidis \etal\ 1995; Taylor \etal\ 1994) and VLBA Calibrator
(VCS; Peck \& Beasley 1998) Surveys.  The sources in the VCS were
primarily selected from the Jodrell Bank-VLA Astrometric Survey (JVAS;
Patnaik \etal\ 1992).

The sources in the COINS sample are described in Table 1. Column (1)
lists the J2000 convention source name of the CSO candidate.  Column
(2) provides an alternate name, with those prefaced by PR or CJ
indicating selection from that survey.  Columns (3) and (4) list the
RA and Declination of the source in J2000 coordinates, and column (5)
shows the optical identification of the source.  

The 19 sources chosen from the PR and CJ surveys were selected based on
criteria described in Readhead \etal\ (1996a) and in Taylor \etal\
(1996a). In brief, an initial selection of objects was made having (1) at 5 GHz
a nearly equal double structure or (2) extended emission on either
side of a strong unresolved component.  These identifications are 
reasonably secure although it is possible further observations could
eliminate one or two sources.
The remaining 33 candidates were selected
from the VCS, and the follow-up observations are described below.  The
VCS is an on-going project to image $\sim$2000 sources at 2.7 and 8.4 GHz
for use as VLBI phase-reference calibrators.  These sources, (in the
declination $-$30\arcdeg\ to +90\arcdeg\ range) have been observed for
$\sim$3 minutes at both frequencies, and the images are available on
the World Wide Web at http://magnolia.nrao.edu/vlba\_calib/.  The
sources in this study were selected from images of the 1500 positive
declination sources in the VCS.  The CSO candidates were identified
based on at least one of the following criteria: a) double structure
at 2.7 GHz, 8.4 GHz or both, where ``double structure'' is considered to
mean having two distinct components with an intensity ratio $<10:1$; 
b) a strong central component with possible extended structure on
both sides at one or both frequencies; c) possible edge-brightening of
one or more components.  

\section{Observations and Analysis}

Observations of the sources selected from the PR and CJ surveys are
reported in Readhead \etal\ (1996) and in Taylor \etal\ (1996a,1996b).
The 33 remaining candidates listed in Table 1 which were
chosen from the VCS were observed as described here along with 
1 candidate CSO from the CJ survey.

The followup observations were made with the 10 element VLBA and a
single VLA antenna, and consisted of 2 observations of 24 hours each.
The first of these took place on 1997 December 28 (1997.99) when 22
sources were observed for 15 minutes each at 8.4 GHz, and 10 of these
having a peak flux $>$100 mJy at 8 GHz in the VCS were also
observed at 15 GHz for $\sim$55 minutes each.  The time spent on each
source was divided into 7 scans which were spread out in hour angle to
obtain good ({\it u,v}) coverage.  The second observation took place
on 1998 March 16 (1998.21).  The remaining 12 candidates were observed
for 15 minutes each at 8.4 GHz, and 17 candidates were also observed
at 5 GHz for $\sim$25 minutes each.  In addition, 3 weak sources,
(having peak fluxes at 8 GHz $<$40 mJy in the VCS), as well as one
source which was found to be very weak at 15 GHz and one which was
found to have extended emission, were observed at 1.6 GHz for $\sim$25
minutes each.

At all frequencies, right and left circular polarizations were
recorded using 2 bit sampling with a total bandwidth of 8 MHz.
Amplitude calibration was derived using the measurements of antenna
gain and system temperature recorded at each antenna, and refined
using the calibrator 3C\,279 (1253$-$055).  Global fringe fitting was
performed using the AIPS task FRING with the following solution
intervals; 7 minutes (for the 15 GHz data), 6 minutes (8 GHz \& 5
GHz), and 5 minutes (1.6 GHz).  In both sets of observations, the
versatile calibrator source 3C\,279 was also used for bandpass
calibration and polarization calibration.  Following the application
of all calibration solutions in AIPS, the data were averaged in
frequency to a single channel.  Editing, imaging and deconvolution
were then performed using {\caps Difmap}~(Shepherd, Pearson, \& Taylor
1995).  Details of each image, including the restoring beam, peak, and
rms noise, are given in Tables 2 and 4.  Flux densities of the various
components in each source were estimated by fitting elliptical
Gaussian models to the self-calibrated visibility data using {\caps
Difmap}. 

\section{Results}

Positive identification of CSOs is contingent upon acquiring
multi-frequency observations in order to correctly identify the core
of the source, which is expected to have a strongly inverted spectrum
(Taylor et al.\ 1996a).  This eliminates any asymmetric core-jet
sources in which a compact jet component might appear similar to the
core component at the discovery frequency.  We have attempted to 
pinpoint the location of the core component for each source using
the criteria of (1) a flat or inverted spectrum; (2) compactness;
and (3) low fractional polarization.    When
extended emission (jets, hot-spots or lobes) are found on both sides
of the core we classify the object as a CSO.  When
the core component is found to be at an extreme end of the source it
is rejected as a CSO, and when no core component can be reliably
identified the source is retained as a candidate CSO.

Images of the sources selected from the VCS which are deemed CSOs or
CSO candidates are shown in Figure 1. Where possible, the core
component has been identified with a cross.  The frequency of the
image shown is indicated in the upper right corner of each plot, and
the beam is displayed in the lower left.  Image parameters are
outlined in Table 2.  The last column in Table 2 indicates which
sources can most reliably be identified as CSOs.  The spectral index
distributions for six of the newly identified CSOs are shown in Figure
2.

Table 3 lists the flux densities of each component at all frequencies
at which follow-up observations were made.  Column (1) lists the J2000
convention source name of the CSO candidate.  Column (2) lists each
component as identified in Figure 1.  Columns (3) through (6) indicate
the total flux density of each component, in mJy/beam, at 1.6, 5, 8.4,
and 15 GHz respectively.  The spectral indices for each component
between 5 and 8.4 GHz are shown in column (7) and those between 8.4
and 15 GHz in column (8).  Spectral indices between 1.6 GHz and other
available frequencies were not calculated because the large
differences in angular resolution and ($\it u,v$) coverage would
render the results unreliable. Upper limits on polarized flux density at 8.4
GHz are shown (in mJy/beam) in column (9).

Images of the sources which have been determined not to be CSOs are
shown in Figure 3.  The frequency of the image shown is indicated in
the upper right corner of each plot, and the beam is displayed in the
lower left.  Image parameters are outlined in Table 4.  The flux
densities of the components are detailed in Table 5, which is
organized in the same manner as Table 3 above.  Column (9) indicates
the peak polarized flux for source components in which this was
measurable, and gives an upper limit for the rest.  Images of the
polarized flux density in the 5 sources with significant detections
are shown in Figure 4.  The polarized flux is shown in grayscale, with
contours from the 8 GHz continuum image superposed.

%

%
%
%
%
%
%
%
%
%
%
%
%
%
%
%
%
%

\section{Discussion}

\subsection{The Incidence of CSOs}

In the surveys we find the incidence of CSOs is 7/65 (11\%) in PR,
18/411 (4.4\%) in PR+CJ, and $\le$39/1900 (2.1\%) in PR+CJ+VCS.  The main
difference between these samples is the parent sample flux limit at 5
GHz which goes from 1.3 Jy in PR, to 0.35 Jy in CJ to $\sim$100 mJy in
the VCS. Although the parent VLBI samples have somewhat different
selection criteria, making it difficult to assess the significance of
the difference in incidence, there does appear to be a trend to a
lower CSO incidence among fainter sources.  Complicating matters
further, however, is the fact that since both the $(u,v)$ coverage and
sensitivity of the VCS are considerably worse than the PR and CJ
surveys, it is possible that some CSOs have been missed in the VCS.
Data quality in the CJ survey was generally better than that from the
earlier PR survey since many more telescopes were available.

%

\subsection{Depolarization by the circumnuclear torus} 

Despite the fact that synchrotron emission is intrinsically polarized
up to 70\% (Burn 1966), less than 0.5\% fractional polarization is
seen in low resolution studies of CSOs at frequencies up to 5 GHz
(Pearson \& Readhead 1988, Taylor \etal\ 1996b).  Even going to high
resolution, Cawthorne \etal\ (1993) found less than 4 mJy of polarized
flux (non-detections) at 5 GHz in the PR CSOs J0111+3906 and
J2355+4950.  In Table 3 we present limits on the polarized flux
density at 8.4 GHz and $\sim$1 mas resolution for 21 CSOs and
candidates in the COINS sample.  In general our 3$\sigma$ limits on
polarized flux density are less than 1.2 mJy/beam.  These correspond
to typical limits on the fractional polarization of $<$1\%, and in
stronger components to as low as $<$0.3\%.  These results are in sharp
contrast to the fractional polarization of 1--20\% typically seen in
the jets of most compact sources (see Cawthorne \etal\ 1993 and our
Table 5).

One possible explanation for the low observed linear polarization from
CSOs is that their radiation is depolarized as it passes through a
magnetized plasma associated with the circumnuclear torus.  In order
to depolarize the radio emission within our 8 MHz IF at 8.4 GHz
the Faraday rotation measures could be larger than 5 $\times$ 10$^5$
radians m$^{-2}$, or alternatively the magnetic fields in the torus
could be tangled on scales smaller than the telescope beam of $\sim$1 mas
to produce gradients of 1000 radian m$^{-2}$ mas$^{-1}$ or
more.

\section{Summary}

From the 33 sources initially selected from the VLBA Calibrator
Survey, we find 10 sources which we can securely classify as Compact
Symmetric Objects.  Thirteen sources, including one previously
unconfirmed candidate from the CJ sample, have been ruled out based on
morphology, spectral index and polarization.  Eleven of the original
VCS sources require further investigation.

Once the redshifts of the remaining newly identified CSOs can be
determined, extensive high spatial and spectral resolution studies of
the neutral hydrogen, as well as the radio continuum and ionized gas
distribution, will be undertaken.  A complete sample of such sources
will yield unique information about accretion processes and the
fueling mechanism by which these young radio galaxies might evolve
into much larger FRII type sources.  Future observations of CSOs
identified in the COINS sample will also be used to measure
bi-directional motions and employ them as cosmological probes.
Studies of the hot spot advance speeds in the COINS sample should
eventually yield kinematic age estimates for the sources and further
advance our understanding of the evolution of radio galaxies.

Finally, it is worth mentioning that CSOs are often useful
calibrators since they are compact, fairly constant in 
total flux density, and have very low polarized flux density.
  
\begin{acknowledgements}

The authors thank Miller Goss for illuminating discussions and
encouragement in the initial stages of this project.  The National
Radio Astronomy Observatory is a facility of the National Science
Foundation operated under a cooperative agreement by Associated
Universities, Inc.  AP is grateful for support from NRAO through the
pre-doctoral fellowship program.  AP acknowledges the New Mexico Space
Grant Consortium for partial publication costs.  This research has
made use of the NASA/IPAC Extragalactic Database (NED) which is
operated by the Jet Propulsion Laboratory, California Institute of
Technology, under contract with the National Aeronautics and Space
Administration.

\end{acknowledgements}

\clearpage


\begin{figure}
\vspace{19cm}
\includegraphics{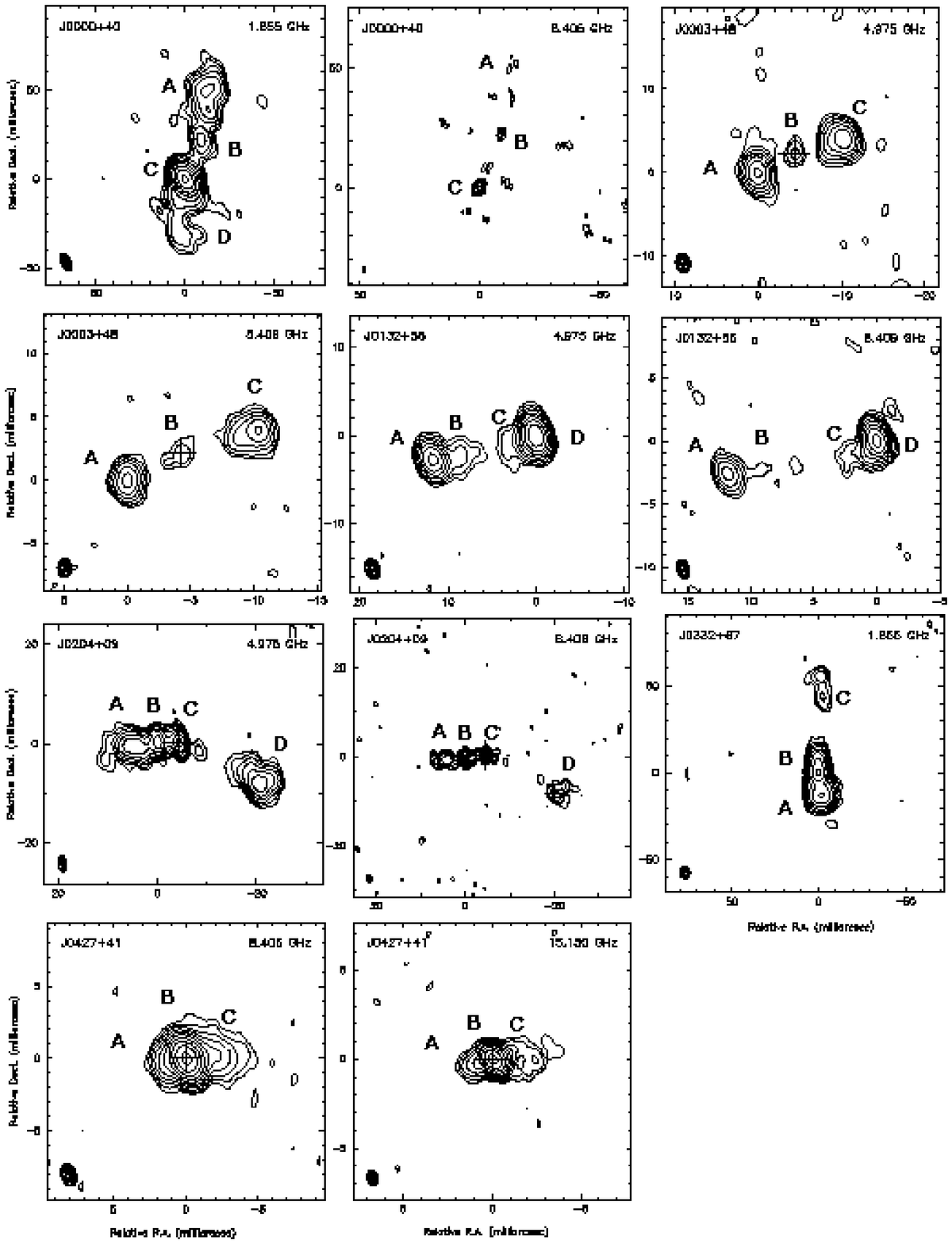}
\figcaption{Continuum images at 1.6, 5.0, 8.4, and 15 GHz, where
available, for the CSO candidates in the COINS sample.  The core
component, when it is possible to identify it as discussed in the
text, is marked with a cross.   The observing
frequency is indicated in the upper right corner of each panel, and
the beam is shown in the lower left.  Image parameters are shown in
Table 2.
\label{fig1}}
\end{figure}

\begin{figure}
\vspace{17cm}
\includegraphics{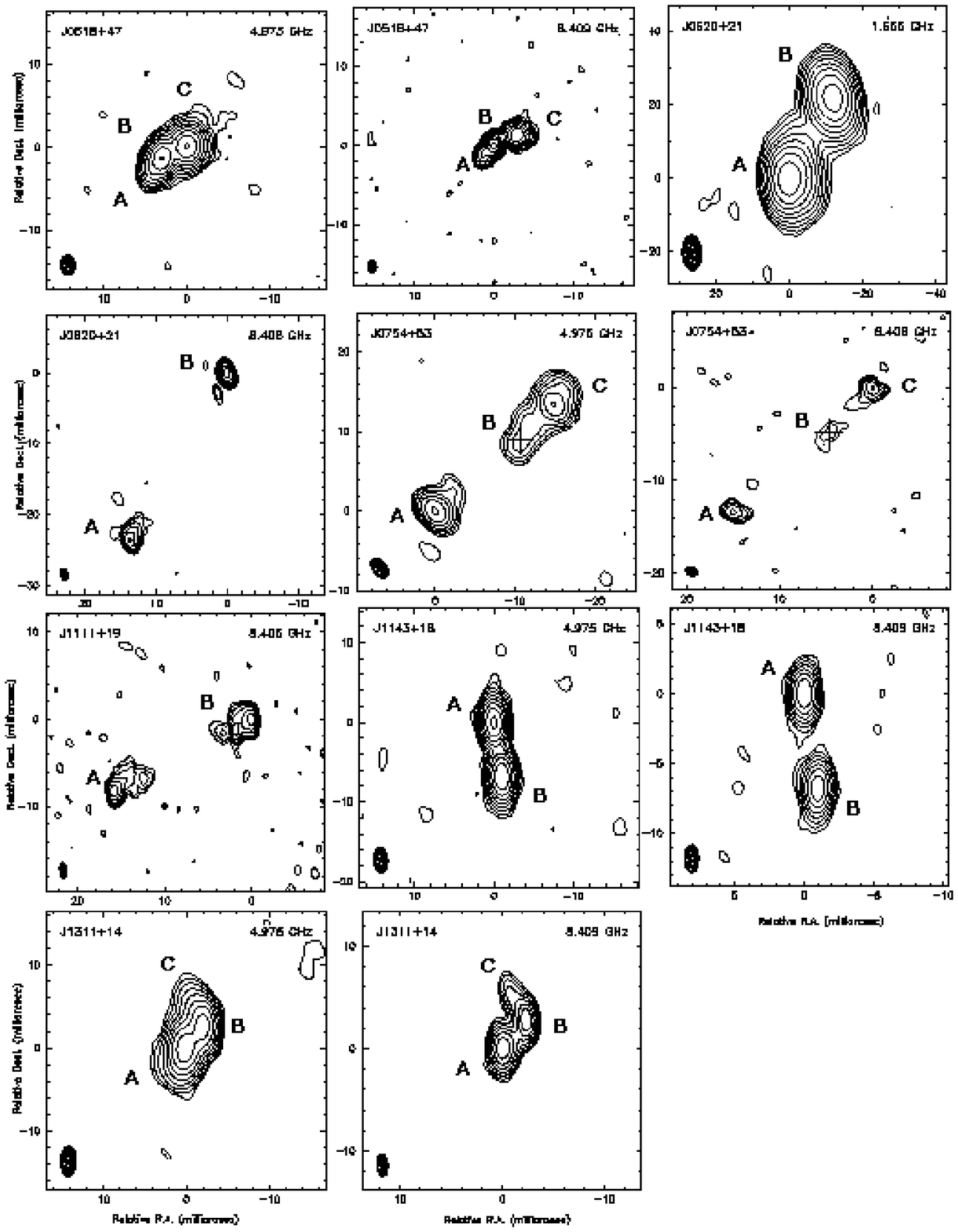}
\figcaption{\caps{CSO candidates -- continued}}
\end{figure}

\begin{figure}
\vspace{17cm}
\includegraphics{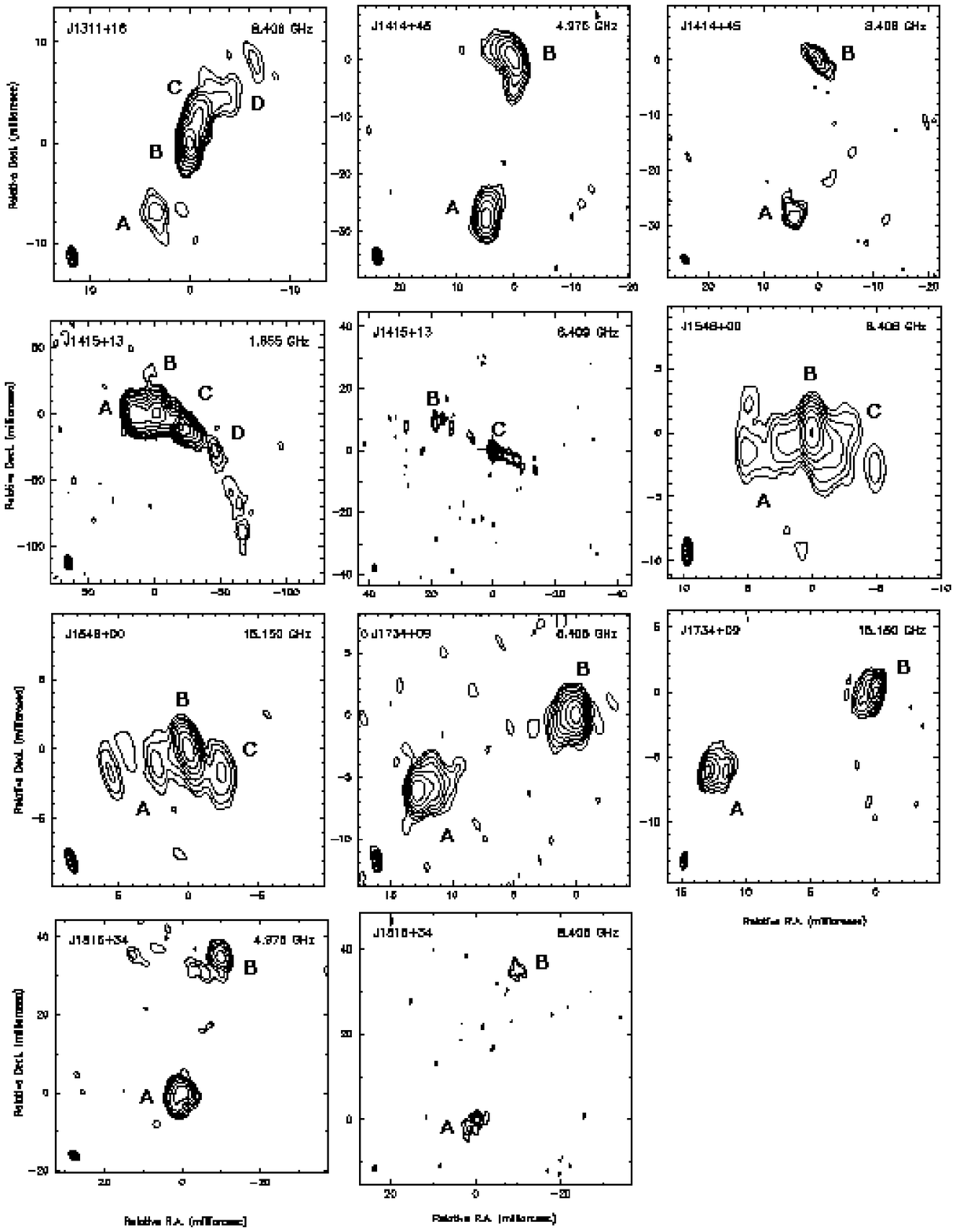}
\figcaption{\caps{CSO candidates -- continued}}
\end{figure}

\begin{figure}
\vspace{17cm}
\includegraphics{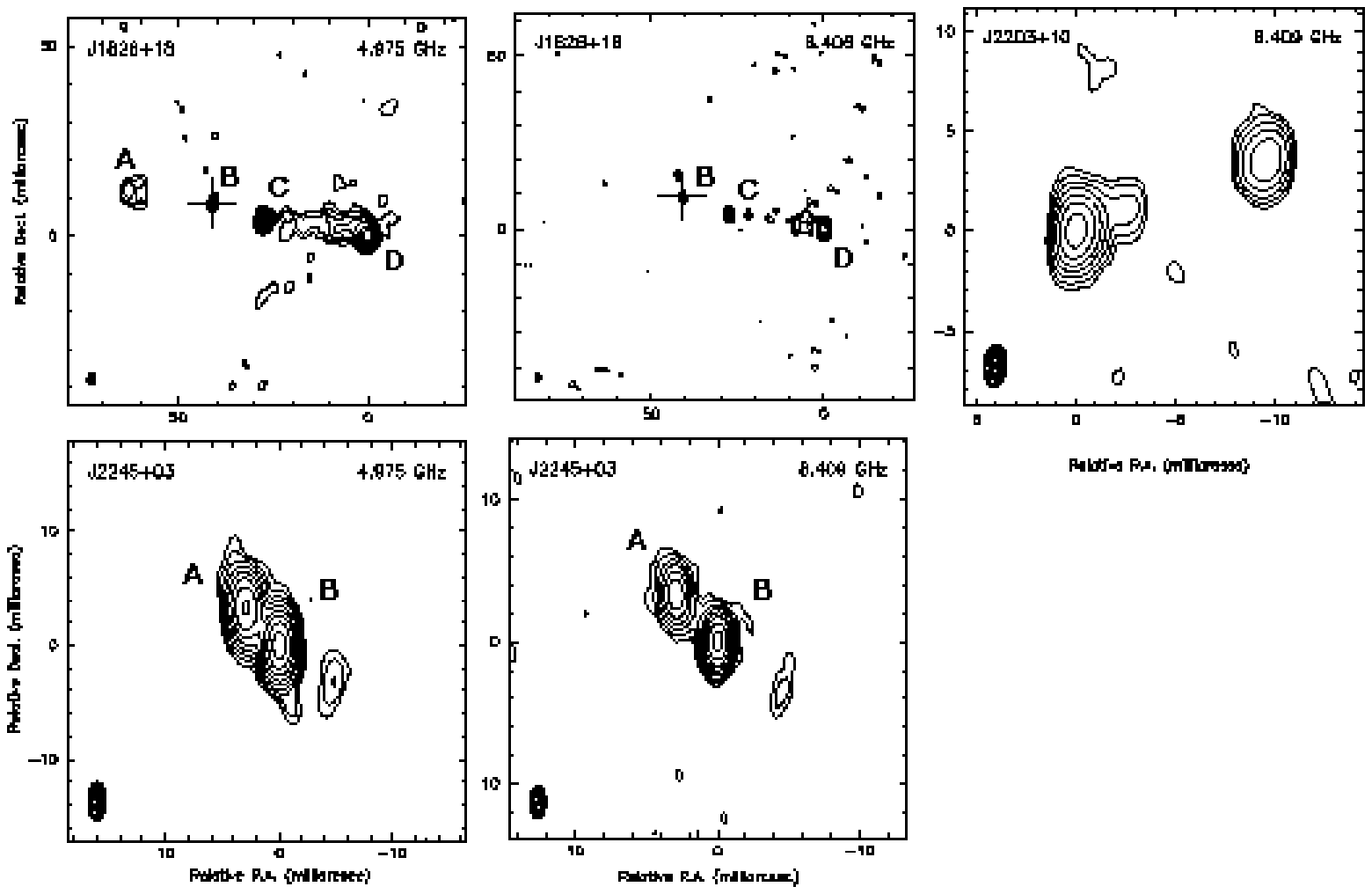}
\figcaption{\caps{CSO candidates -- continued}}
\end{figure}

\begin{figure}
\vspace{16cm}
\includegraphics{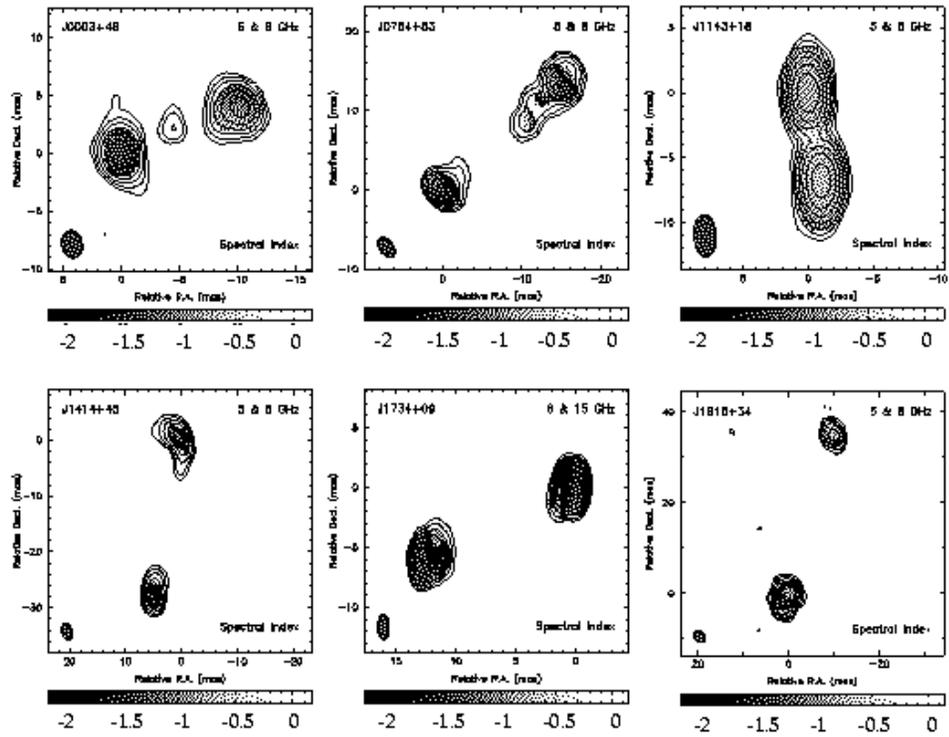}
\figcaption{Spectral index distributions of 6 of the confirmed CSOs in
the COINS sample.  The frequencies used to calculate the spectral
indices are shown in the top right corner of each panel, and the
contours shown are from the continuum image at the lower frequency for
each source.
\label{fig2}}
\end{figure}

\begin{figure}
\vspace{19cm}
\includegraphics{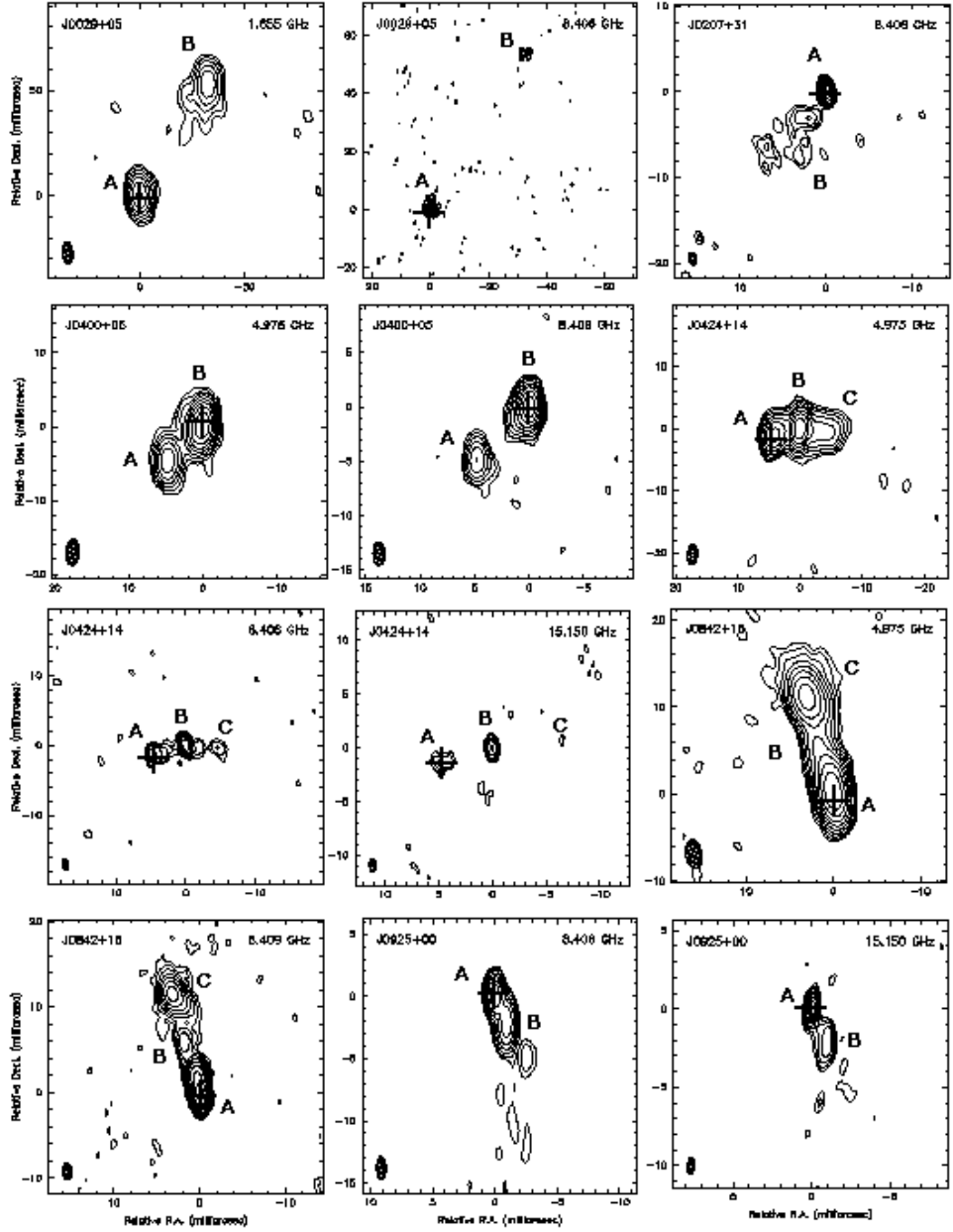}
\figcaption{Continuum images at 1.6, 5.0, 8.4, and 15 GHz, where
  available, for the sources which are no longer considered CSO
  candidates.  The core component is marked with a cross.  The
  observing frequency is indicated in the upper right corner of each
  panel, and the beam is shown in the lower left.  Image parameters
  are shown in Table 4.
\label{fig3}}
\end{figure}

\begin{figure}
\vspace{17cm}
\includegraphics{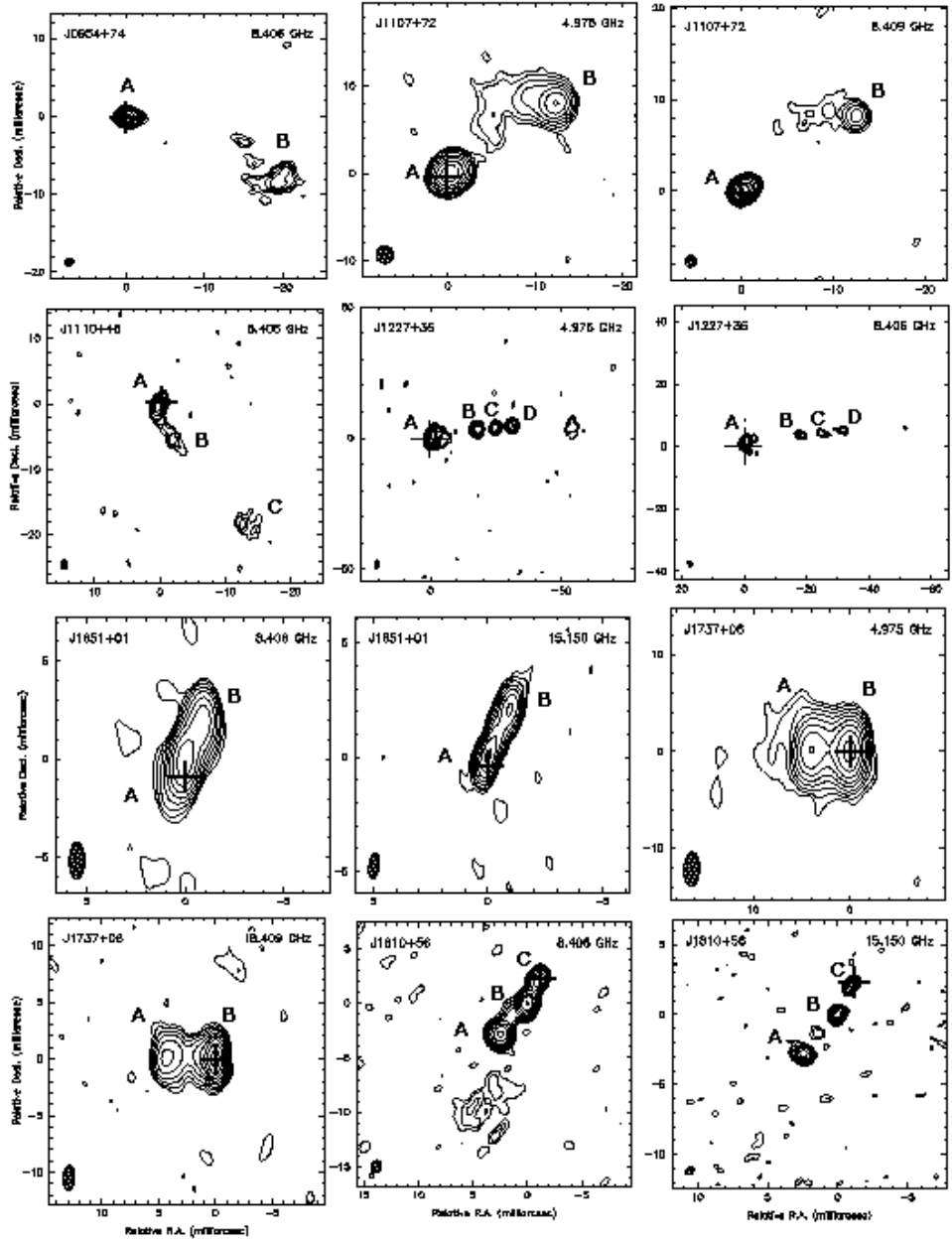}
\figcaption{\caps {Rejected Sources -- Continued}}
\end{figure}

\begin{figure}
\vspace{16cm}
\includegraphics{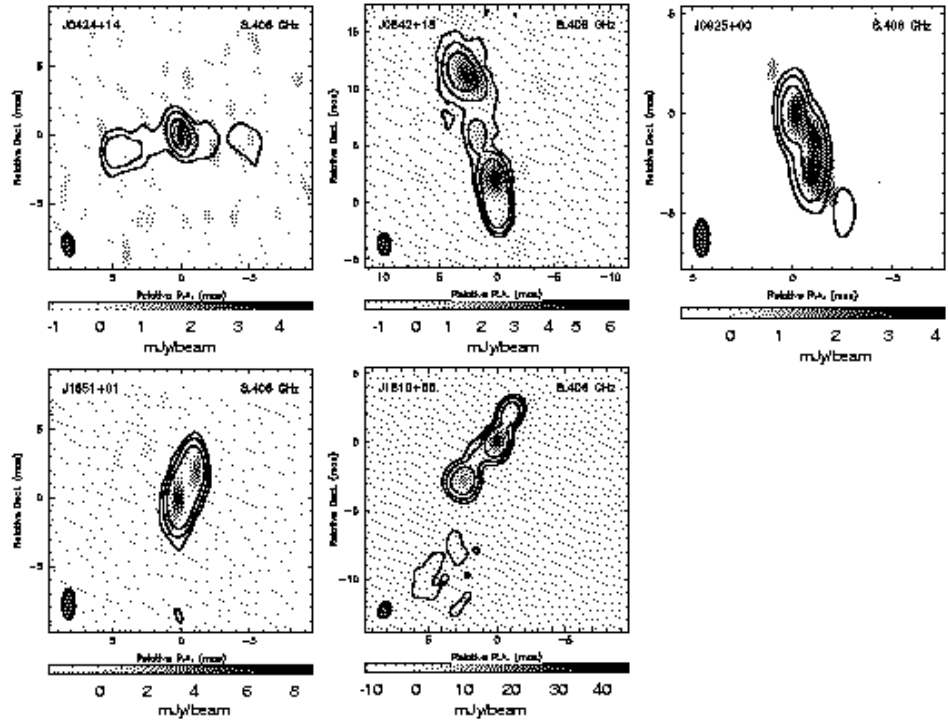}
\figcaption{Images of polarized flux in the five sources in the COINS
sample in which polarization is detected.  Contours shown are from the
8.4 GHz continuum images.
\label{fig4}}
\end{figure}

\clearpage

\renewcommand{\baselinestretch}{1.2}
\scriptsize
\begin{table}[h]
\begin{center}
\caption{The complete list of sources in the COINS sample\label{tab1}}
~~~\\
\begin{tabular}{llrrc}
\tableline\tableline
Source &Alternate& &     \\
Name &Name &{RA} &{Dec} & ID   \\
(1) &(2) &(3) &(4)  &(5)   \\
\tableline
J0000+4054 &4C 40.52&00 00 53.0815&40 54 01.806&G   \\
J0003+4807 &&00 03 46.0413&48 07 04.134&... \\
J0029+3456&CJ 0026+346&00 29 14.2436&34 56 32.255 &G  \\
J0029+0509& &00 29 03.5922&05 09 34.860&QSO \\
J0111+3906&PR 0108+388&01 11 37.3192&39 06 28.085&G \\
J0132+5620& &01 32 20.4503&56 20 40.372& ... \\
J0204+0903& &02 04 34.7589&09 03 49.248&...  \\
J0207+3152&5C 06.008&02 07 34.9902&31 52 06.458&G \\
J0332+6753& &03 32 59.5241&67 53 03.860&... \\
J0400+0550& &04 00 11.7358&05 50 43.135&... \\
J0410+7656&PR 0404+768&04 10 45.6057&76 56 45.301 &G \\
J0424+1442& &04 24 23.4914&14 42 16.688&...  \\
J0427+4133& &04 27 46.0455&41 33 01.091&...  \\
J0518+4730& &05 18 12.0899&47 30 55.536&...  \\
J0620+2102& &06 20 19.5286&21 02 29.501&...  \\
J0650+6001&CJ 0646+600&06 50 31.2556&60 01 44.547&QSO \\
J0713+4349&PR 0710+439&07 13 38.1642&43 49 17.199&G \\
J0753+4231&CJ 0749+426&07 53 03.3378&42 31 30.763&QSO  \\
J0754+5324 &&07 54 15.2177&53 24 56.450&...  \\
J0842+1835 &&08 42 05.0944&18 35 40.987&QSO  \\
J0925+0019& &09 25 07.8153&00 19 13.929&QSO  \\
J0954+7435 &&09 54 47.4440&74 35 57.140&G   \\
J1035+5628&PR 1031+567&10 35 07.0399&56 28 46.792&G \\
J1107+7232& &11 07 41.7240&72 32 36.003&QSO \\
J1110+4817&&11 10 36.3237&48 17 52.446&QSO  \\
J1111+1955& &11 11 20.0694&19 55 35.950&G  \\
J1143+1834& &11 43 26.0706&18 34 38.375&...  \\
J1148+5924&CJ 1146+596&11 48 50.3670&59 24 56.380&G \\
J1227+3635& &12 27 58.7260&36 35 11.819&QSO  \\
J1244+4048&CJ 1242+410&12 44 49.1879&40 48 06.137&QSO \\
J1311+1417& &13 11 07.8250&14 17 46.659&QSO  \\
J1311+1658& &13 11 23.8204&16 58 44.213&...  \\
J1357+4353&CJ 1355+441&13 57 40.6762&43 53 59.671&G \\
J1400+6210&PR 1358+624&14 00 28.6526&62 10 38.526&G \\
J1414+4554& &14 14 14.8526&45 54 48.730&G   \\
J1415+1320&PKS1413+135&14 15 58.8188&13 20 23.714&QSO  \\
J1546+0026& &15 46 09.5312&00 26 24.615&G \\
\end{tabular}
\end{center}
\end{table}

\clearpage
\renewcommand{\baselinestretch}{1.2}
\scriptsize
\begin{table}[h]
\begin{center}
TABLE 1. Continued
\vskip 2pt
\begin{tabular}{llrrc}
\tableline\tableline
Source & Alternate &   &     \\
Name &Name &{RA} &{Dec} & ID   \\
(1) &(2) &(3) &(4)  &(5)  \\
\tableline
J1651+0129& &16 51 03.6620&01 29 23.448&...  \\
J1734+0926& &17 34 58.3773&09 26 58.274&G    \\
J1737+0621& &17 37 13.7290&06 21 03.567&QSO  \\
J1810+5649&CJ 1809+568&18 10 03.3203&56 49 22.959&... \\
J1815+6127&CJ 1815+614&18 15 36.7920&61 27 11.641&QSO \\
J1816+3457& &18 16 23.8987&34 57 45.729&G  \\
J1826+1831& &18 26 17.7118&18 31 52.915&... \\
J1823+7938&CJ 1826+796&18 23 14.1088&79 38 49.002&G \\
J1845+3541&CJ 1843+356&18 45 35.1097&35 41 16.719&G  \\
J1944+5448&CJ 1943+546&19 44 31.5138&54 48 07.069&G \\
J1945+7055&CJ 1946+708&19 45 53.5197&70 55 48.723&G \\
J2022+6136&PR 2021+614&20 22 06.6820&61 36 58.806&QSO \\
J2203+1007 &&22 03 30.9534&10 07 42.584&... \\
J2245+0324 &&22 45 28.2846&03 24 08.863&QSO  \\
J2355+4950&PR 2352+495&23 55 09.4582&49 50 08.340&G \\
\tableline
\end{tabular}
\end{center}
\end{table}

\begin{table}
\begin{center}
\caption{CSO Candidate Image Parameters\label{tab2}}
~~~\\
\begin{tabular}{lrcrrccc}
\tableline\tableline
       &       &       &         &Peak  Flux &  rms & Lowest & \\
Source & Freq. &  Beam & $\theta$ &(mJy &(mJy  & Contour & Status \\
Name   & (GHz) & (mas) & & beam$^{-1}$) &  beam$^{-1}$) & (mJy)
&  \\

\tableline
J0000+4054 &1.6 &10.1$\times$5.38&29.7&436.5&0.5& 1.5 & CAND \\
           &8.4 &2.32$\times$1.14& 5.4&59.2 &0.4& 1.5 & \\
J0003+4807 &4.9 &2.39$\times$1.75&14.1&60.9 &0.4& 0.8 & CSO\\
           &8.4 &1.51$\times$1.09& 1.8&24.8 &0.3& 0.6 & \\

J0132+5620 &4.9 &2.37$\times$1.58&18.6&376.0&0.5& 1.5 & CAND \\
           &8.4 &1.55$\times$1.01&18.1&203.0&0.5& 1.2 & \\
J0204+0903 &4.9 &3.24$\times$1.68& 7.3&97.2 &0.3& 1.0 & CSO \\
           &8.4 &1.82$\times$0.95& 5.6&54.0 &0.5& 1.5 & \\

J0332+6753 &1.6 &7.32$\times$6.18& 6.3&114.0&0.3& 0.8 & CAND \\

J0427+4133 &8.4 &1.63$\times$1.01&18.2&494.8&0.3& 0.8 & CSO\\
           &15.0&0.86$\times$0.59&12.8&269.9&0.3& 0.8 & \\
J0518+4730 &4.9 &2.49$\times$1.68&0.07&215.1&0.3& 0.8 & CAND  \\
           &8.4 &1.56$\times$1.06&-0.5&126.2&0.3& 0.8 & \\
J0620+2102 &1.6 &9.60$\times$5.35&2.2 &367.4&0.3& 0.9 & CAND\\
           &8.4 &1.72$\times$1.00&22.3&117.6&0.4& 1.2 & \\
J0754+5324 &4.9 &2.80$\times$1.68&36.5&56.7 &0.3& 0.8 & CSO\\
           &8.4 &1.32$\times$0.93&65.1&17.9 &0.4& 1.0 & \\

J1111+1955 &8.4 &1.73$\times$0.77&5.5 &50.6&0.5& 1.0 & CAND\\
%
J1143+1834 &4.9 &3.26$\times$1.71&3.1 &177.7&0.2& 0.6 &CAND \\
           &8.4 &1.98$\times$1.05&0.5 &114.8&0.3& 0.9 & \\

J1311+1417 &4.9 &3.61$\times$1.85&$-$0.9&170.0&0.4& 0.8 & CAND  \\
           &8.4 &2.15$\times$1.06&3.7 &92.5 &0.2& 0.8 & \\
J1311+1658 &8.4 &2.15$\times$1.01&8.2 &137.9&0.3& 0.8 & CAND  \\
J1414+4554 &4.9 &3.01$\times$1.77&13.5&39.0 &0.3& 0.8 & CSO\\
           &8.4 &1.84$\times$1.30&37.5&15.9 &0.3& 0.8 & \\
J1415+1320 &1.6 &10.9$\times$5.79&7.4 &147.9&0.3& 0.9 &CSO \\
           &8.4 &2.26$\times$1.12&4.9 &1241.3&0.6& 2.0& \\
J1546+0026 &8.4 &2.38$\times$0.98&0.1 &289.4&0.6& 2.0 & CSO\\
           &15.0&1.86$\times$0.72&16.9&193.7&0.6& 1.8 & \\

J1734+0926 &8.4 &2.11$\times$0.95&1.1 &160.8&0.2& 0.8 & CSO\\
           &15.0&1.38$\times$0.55&$-$7.2&65.5 &0.3& 0.8 & \\

J1816+3457 &4.9 &2.87$\times$2.21&43.1&75.2 &0.4 &1.2 & CSO\\
           &8.4 &1.55$\times$0.82&$-$5.9&21.3 &0.4 &1.4 & \\

J1826+1831 &1.6 &1.84$\times$3.12&$-$7.6&116.0&0.4&1.0 & CSO \\
           &8.4 &1.10$\times$2.25&4.2&51.3&0.4&1.0 & \\

J2203+1007 &8.4 &1.12$\times$2.14&$-$5.1&95.7&0.4&1.0  & CAND \\
J2245+0324 &4.9 &3.34$\times$1.58&0.13&357.0&0.4&1.0 & CAND  \\
           &8.4 &1.10$\times$2.24&$-$2.0&259.9&0.6&1.4 & \\

\tableline
\end{tabular}
\end{center}
\tablenum{2}
\end{table}
\clearpage

\renewcommand{\baselinestretch}{1.2}
\scriptsize
\begin{table}
\tablenum{3}
\begin{center}
\caption{CSO Candidates from the VLBA Calibrator Survey\label{tab3}}
~~~\\
\begin{tabular}{lcrrrrrrr}
\tableline\tableline
Source &Comp. &$S_{1.6}$   &$S_5$ &$S_{8.4}$ &$S_{15}$ &$\alpha^{5}_{8.4}$
 &$\alpha^{8.4}_{15}$ &P$_{8.4}$ \\
(1) &(2) & (3)  &(4) &(5) &(6) &(7) &(8) &(9)\\
\tableline
J0000+4054 &A& 228.9  &... &28.7 &... &...  &... &$<$1.2\\
 &B & 47.3&... &35.2 &... &... &... & \\
 &C & 688.5&... &173.7 &... &... &...& \\
 &D & 32.2&... &... &... &... &...& \\
J0003+4807 &A& ... & 70.6 &36.9 &... &$-$1.24  &...  &$<$0.9\\
 &B     &  ...&  3.5&3.4 &... & $-$0.06&...& \\
 &C &... & 66.0& 40.3& ...& $-$0.94&... & \\

J0132+5620 &A&... &140.8&34.6&... &$-$2.68&...&$<$0.9 \\
 &B &... &20.1 & 4.7&... &$-$2.77 &...& \\
 &C &... &7.0 & 6.5&... &$-$0.14 &... & \\
 &D & ...& 436.1& 245.3&... &$-$1.10 &... &\\
J0204+0903 &A&... &181.3&103.5&...&$-$1.07&...&$<$1.2 \\
 &B &... &122.6 & 75.6&... &$-$0.92 &... & \\
 &C &... & 90.1& 81.2&... &$-$0.20 &...& \\
 &D &.. & 155.6& 57.1&... &$-$1.91 &... & \\

J0332+6753 &A&95.8&...&...&...&...&...& \\ 
 &B &207.3 &... &... &... &... &...& \\
 &C &10.9 &... &... & ...& ...&...& \\

J0427+4133 &A&... &...&32.7&22.0&... &$-$1.46&$<$0.9\\
 &B &... &... &588.9 &406.0 &... &$-$0.63& \\
 &C &... &... &28.7 &13.9 &... &$-$1.23 & \\

J0518+4730 &A&...&56.1&35.7&...&$-$0.86&...&$<$0.7 \\
 &B &... &217.1 &134.2 &... &$-$0.92 &...&  \\
 &C &... &304.5 &174.9 &... &$-$1.06 &...&  \\
J0620+2102 &A&478.4&...&77.2&... &... &...&$<$0.9 \\
 &B &400.0 &... &148.4 &... &... &...& \\
J0754+5324 &A&...&76.1&29.4&... &$-$1.81&...&$<$1.2 \\
 &B &... &27.6 &10.2 &... &$-$0.52 &... & \\
 &C &... &79.6 &34.2 &... &$-$1.61 &... &\\

J1111+1955 &A&...&...&98.2&...&...&...&$<$0.9 \\
 &B &... &... &126.5 &... &... & ...& \\
J1143+1834 &A&...&180.7&121.2&... &$-$0.76&... &$<$0.6 \\ 
&B &... &159.8 &112.7 &... &$-$0.67 & ...& \\

J1311+1417 &A&...&214.3&127.1&...&$-$1.00&...&$<$0.9 \\
 &B &... &148.5 &103.0 &... & $-$0.70&...& \\
 &C &... &33.6 &20.8 &... &$-$0.91 &...&\\
J1311+1658 &A&...&...&8.6 &...&...&...&$<$1.2\\
 &B &... &... &208.7 &... &... &...& \\
 &C &... &... &11.4 &... &... &...& \\
 &D &... &... &15.6 &... &... &... &\\

\end{tabular}
\end{center}
\tablenum{3}
\end{table}
\clearpage
\renewcommand{\baselinestretch}{1.2}
\scriptsize
\begin{table}[h]
\begin{center}
TABLE 3. Continued
\vskip 2pt
\begin{tabular}{lcrrrrrrr}
\tableline\tableline
Source &Comp. &$S_{1.6}$   &$S_5$ &$S_{8.4}$ &$S_{15}$ &$\alpha^{5}_{8.4}$
 &$\alpha^{8.4}_{15}$&P$_{8.4}$ \\
(1) &(2) & (3)  &(4) &(5) &(6) &(7) &(8) &(9) \\
\tableline

J1414+4554 &A&...&76.1&34.2&... &$-$1.52&...&$<$1.2 \\
 &B &... &97.8 &41.9 & ...&$-$1.62 &...& \\

J1415+1320 &A&260.9&... &...  &... &... &...&$<$4.5\\
 &B &390.8  &...&44.6 &... &... &... &\\
 &C &250.2 &... &1381.8 &... &... &... & \\
 &D &21.8 &... &... & ...&... & ... &\\

J1546+0026 &A&...&...&94.2&60.7&... &$-$0.75&$<$0.9 \\
 &B &... &... &354.0 &242.9 & ...&$-$0.64& \\
 &C &... &... &87.8 &71.7 & ...&$-$0.34& \\

J1734+0926 &A&...&...&162.4&66.7&... &$-$1.51&$<$1.2 \\
 &B &... &... &238.8 &97.2 &... &$-$1.53 & \\

J1816+3457 &A&...&196.8&74.4&... &$-$1.85&... &$<$1.2 \\
 &B & ...&75.9 &27.7 &... &$-$1.92 &... & \\

J1826+1831 &A&...&16.0&...&...&...&...&$<$1.5 \\
 &B & ... &5.3&8.3 &... &0.86 &... & \\
 &C &... &50.4 &31.0 &... &$-$0.93 &... & \\
 &D &... &200.7 &107.2 &... &$-$1.20 &... & \\

J2203+1007 &A&...&...&120.9&...&...&...&$<$1.2 \\
 &B & ...&... &7.8 &... &... &... & \\
 &C &... &... &44.0 &... &... &... &\\
J2245+0324 &A&...&166.8 &116.9&...&$-$0.68&...&$<$1.2 \\
 &B &... &369.2 &267.2&... &$-$0.62 &... &\\

\tableline
\end{tabular}
\end{center}
\end{table}
\clearpage

\begin{table}
\tablenum{4}
\begin{center}
\caption{Core-Jet Source Image Parameters\label{tab4}}
~~~\\
\begin{tabular}{lrcrrrrr}
\tableline\tableline
       &       &       &         &Peak  Flux &  rms & Lowest & \\
Source & Freq. &  Beam & $\theta$ &(mJy &(mJy  & Contour & \\
Name   & (GHz) & (mas) & & beam$^{-1}$) &  beam$^{-1}$) & (mJy) \\
\tableline
J0029+0509 &1.6 &10.8$\times$4.86& 5.1&110.0&0.5& 1.5 & \\
           &8.4 &1.99$\times$1.00&$-$2.6&130.0&0.6& 1.5 & \\
J0207+3152 &8.4 &1.53$\times$0.84&11.3&173.0&0.4& 1.2 & \\
J0400+0550 &4.9 &3.52$\times$1.73&$-$2.9&310.8&0.4& 1.2 & \\
           &8.4 &2.15$\times$1.07&0.7 &275.1&0.4& 1.2 & \\
J0424+1442 &4.9 &3.36$\times$1.61&$-$3.8&150.9&0.2& 0.6 & \\
           &8.4 &1.68$\times$0.84&10.4&74.9 &0.4& 1.2 & \\
           &15.0&1.19$\times$0.61&1.9 &20.8 &0.5& 1.2 & \\
J0842+1835 &4.9 &3.37$\times$1.83&8.9 &525.5&0.4& 1.0 & \\
           &8.4 &1.96$\times$1.09&5.2 &394.3&0.4& 1.0 & \\
J0925+0019 &8.4 &1.84$\times$0.76&0.7 &151.7&0.3& 1.2 & \\
           &15.0&1.04$\times$0.45&$-$4.4&151.5&0.4& 1.2 & \\
J0954+7435 &8.4 &1.01$\times$0.89&$-$68.6&100.6&0.4&1.0 & \\
J1107+7232 &4.9 &1.97$\times$1.96&16.5&132.4&0.2&0.6 & \\
           &8.4 &1.25$\times$1.14&$-$32.9&175.6&0.2&0.6 & \\
J1110+4817 &8.4 &1.38$\times$0.92&$-$7.5&32.6&0.4& 1.0 & \\
J1227+3635 &4.9 &2.95$\times$1.70&$-$4.2&250.0&0.6& 1.8 & \\
           &8.4 &1.69$\times$1.06&38.5&79.0 &0.3& 1.4 & \\
J1651+0129 &8.4 &2.23$\times$0.93&$-$0.3&289.7&0.3& 1.2 & \\
           &15.0&1.10$\times$0.43&$-$7.7&178.7&0.4& 1.2 & \\
J1737+0621 &4.9 &3.48$\times$1.60&$-$1.7&277.8&0.3& 0.8 & \\
           &8.4 &2.15$\times$1.00&$-$1.7&436.1&0.5& 1.2 & \\
J1810+5649 &8.4 &1.21$\times$0.89&$-$23.0&221.5&0.4&0.8 & \\
           &15.0&0.61$\times$0.50&$-$45.8&157.7&0.7&1.4 & \\

\tableline
\end{tabular}
\end{center}
\tablenum{4}
\end{table}
\clearpage

\renewcommand{\baselinestretch}{1.2}
\scriptsize
\begin{table}[h]
\tablenum{5}
\begin{center}
\caption{Sources Rejected as CSOs\label{tab5}}
~~~\\
\begin{tabular}{lcrrrrrrr}
\tableline\tableline
Source &Comp. &$S_{1.6}$   &$S_5$ &$S_{8.4}$ &$S_{15}$ &$\alpha^{5}_{8.4}$
 &$\alpha^{8.4}_{15}$ & P$_{8.4}$ \\
(1) &(2) & (3)  &(4) &(5) &(6) &(7) &(8) & (9) \\
\tableline
J0029+0509 &A& 150.3 &...&267.7&... &...  &... &$<$1.2\\
 &B & 103.4&... &50.1 &... &... &... & \\
J0207+3152 &A&... &...&179.1&...&...&...&$<$2.0\\
 &B &... &... &76.7 &... &... &... & \\
J0400+0550 &A&... &80.8 &50.2&...&$-$0.91&...&$<$0.9 \\
 &B &... &360.0 &318.4 &... &$-$0.23 &... & \\
J0424+1442 &A&... &78.6 &43.8&14.2&$-$1.11&$-$1.91&$<$1.1\\
 &B &... &182.9 &100.8 & 28.5&$-$1.14 &$-$2.14 &4.1\\
 &C &... &36.6 &14.6 & 1.2&$-$1.75 &... &$<$1.1 \\
J0842+1835 &A&...&676.3&595.2&...&$-$0.24&...&6.1 \\
 &B &... &71.7 &21.6 & ...&$-$2.29 &... &1.3 \\
 &C &... &281.7 &188.8 &... &$-$0.76 &... &5.3 \\
J0925+0019 &A&...&...&156.7&156.6&...&$-$0.00&3.7 \\
 &B &... &... &124.7 &59.8 &... &$-$1.25 &3.7 \\
J0954+7435 &A&...&...&193.8&...&...&...&$<$1.2 \\
 &B &... &... &151.1 &... &... &... & \\
J1107+7232 &A&... &164.9&210.0&...&0.46&...&$<$0.6 \\
 &B &... &52.9 &32.2 &... &$-$0.95 &... & \\
J1110+4817 &A&... &...&70.7&...&...&...&$<$1.2\\
 &B &... &... &18.7 &... &... &... & \\
 &C &... &... &18.8 &... &... &... & \\
J1227+3635 &A&...&498.9&191.9&...&$-$1.82 &...&$<$1.2 \\
 &B &... &92.6 &29.1 &... &$-$2.21 & ...& \\
 &C & ...&42.9 &16.4 &... &$-$1.83 &... & \\
 &D & ...&49.5 &18.0 &... &$-$1.93 &... &\\
J1651+0129 &A&... &...&331.1&291.1&...&$-$0.22 &8.0\\
 &B &... &... &183.2 &121.4 &... &$-$0.70 & 5.1\\
J1737+0621 &A&...&164.5&107.6&...&$-$0.81&...&$<$1.2  \\
 &B &... &339.5 &500.8&... &0.74 &... &\\
J1810+5649 &A&...&...&159.8&113.0&... &$-$0.59&18.0  \\
 &B &... &... &235.5 &169.4 &... &$-$0.56 &41.7 \\
 &C &... &... &86.0 &67.2 &... &$-$0.42 &5.6 \\
\tableline
\end{tabular}
\end{center}
\end{table}
\clearpage

\end{document}